\documentclass{PoS}
\pdfoutput=1

\usepackage{amssymb}
\usepackage{amsmath}
\usepackage{units,calc}
\usepackage[linesnumbered,algoruled,longend]{algorithm2e}
\usepackage[font=small,labelfont=bf,tableposition=top]{caption}
\usepackage{tabularx}
\usepackage{array}
\usepackage{pgf}

\newcolumntype{Y}{>{\centering\arraybackslash}X}

\newcommand{\SetAlgorithmStyle}{
  \SetKwFor{ForAll}{for all}{do}{end for}%
  \SetKwFor{ForEach}{foreach}{do}{end for}%
  \SetKwFor{For}{for}{do}{end for}%
  \SetArgSty{}
  \DontPrintSemicolon
}
\SetCommentSty{emph}

\newcommand\WMG{\mbox{DD-$\alpha$AMG}}
\newcommand\Ntv{N_\text{tv}}

\newcommand\Nsetup{N_\text{setup}}
\newcommand\Nsimd{N_\text{SIMD}}
\newcommand\Nb{N_\text{b}}
\newcommand\Nblock{N_\text{block}}
\newcommand\qp{QPACE~2}

\DeclareMathOperator{\re}{Re}
\DeclareMathOperator{\im}{Im}

\title{Multiple right-hand-side setup for the \WMG\thanks{Work
    supported by the German Research Foundation (DFG) in the framework
    of SFB/TRR-55.}}

\ShortTitle{MRHS setup for the \WMG}

\author{\speaker{Daniel Richtmann}$^{\,a}$,
        Simon Heybrock$^{a,b}$,
        Tilo Wettig$^a$\\
        $\phantom{}^a$Department of Physics, University of Regensburg, 93040
        Regensburg, Germany\\
        $\phantom{}^b$Data Management and Software Centre, European Spallation
        Source, Universitetsparken 5, 2100 Copenhagen, Denmark\\
        E-mail: \email{daniel.richtmann@ur.de}}

\abstract{
  The setup cost of a modern solver such as \WMG\ (Wuppertal Multigrid) is a
  significant contribution to the total time spent on solving the Dirac
  equation, and in HMC it can even be dominant. We present an improved
  implementation of this algorithm with modified computation order in the setup
  procedure. By processing multiple right-hand sides simultaneously we can
  alleviate many of the performance issues of the default single right-hand-side
  setup. The main improvements are as follows:\\
  By combining multiple right-hand sides the message size for off-chip
  communication is larger, which leads to better utilization of the network
  bandwidth. Many matrix-vector products are replaced by matrix-matrix products,
  leading to better cache reuse. The synchronization overhead inflicted by
  on-chip parallelization (threading), which is becoming crucial on many-core
  architectures such as the Intel Xeon Phi, is effectively reduced. In the parts
  implemented so far, we observe a speedup of roughly 3x compared to the
  optimized version of the single right-hand-side setup on realistic lattices.
}

\FullConference{The 33rd International Symposium on Lattice Field Theory\\
                14 -18 July 2015\\
                Kobe International Conference Center, Kobe, Japan}

\begin{document}
\section{Introduction and motivation}
Conventional iterative Krylov subspace solvers for the Dirac equation share a
common behavior when going to small quark masses: Their iteration number and
time to solution (wall-clock time) increases drastically, which basically
renders them unusable. Therefore a lot of effort has been put into developing
efficient preconditioning algorithms that aim at tackling this problem, such as
domain decomposition \cite{Luscher:2003qa}, inexact deflation
\cite{Luscher:2007se}, and multigrid approaches
\cite{Babich:2010qb,Frommer:2013fsa}. While these methods significantly reduce
the iteration number, the latter two introduce an additional overhead compared
to standard solvers since they require an initial setup phase before one can
start solving the Dirac equation. In HMC the setup cost can even be the dominant
contribution to the total wall-clock time spent in the solver since only a few
solves can be done before the setup has to be updated. Thus an optimization of
the setup code potentially has a large impact on the overall HMC performance.

Based on the attractive theoretical properties and the performance of the \WMG\
algorithm, the Regensburg group (RQCD) recently decided to port the
implementation of this algorithm by the Wuppertal group \cite{Frommer:2013fsa},
which is C-MPI code aimed at standard CPUs, to SIMD architectures, with a
special focus on the Intel Xeon Phi architecture (KNC) used in
\qp~\cite{Arts:2015jia}. This involved threading the code using OpenMP,
optimizing it for the wide SIMD registers of the KNC, and reducing
memory-bandwidth requirements by enabling the use of half precision on the
coarse grid. For a detailed description of this effort see
\cite{Heybrock:lat15}.

Even with the improvements achieved in \cite{Heybrock:lat15}, there is still
optimization potential in the setup of \WMG, as it remains expensive. In this
contribution we document our work on an improved implementation that modifies
the computation order in the setup phase to process multiple right-hand sides
simultaneously.

\section{Description of the algorithm}

The \WMG\ algorithm uses FGMRES as the outer Krylov subspace solver for the
Dirac equation, preconditioned by a multigrid method that consists of two parts:
a smoother working on the fine grid that reduces the error contribution of
eigenvectors with large eigenvalues (high modes), and a coarse-grid correction
(CGC) that reduces the error contribution of low modes.\footnote{As in
\cite{Heybrock:lat15} we restrict ourselves to two grid levels.} To this end
projection operators between the grids and the Dirac operator on the coarse grid
need to be defined in an initial setup phase.

The setup procedure of \WMG\ is based on a set of $\Ntv$ random test vectors
(each of dimension $12V$, where $V$ is the lattice volume) that are used to
construct restriction $R$, prolongation $P = R^{\dagger}$, and coarse-grid
operator $D_c$. The setup is split into two parts: an initial phase and an
iterative refinement phase. In the initial phase, a domain-decomposition (DD)
smoother based on the Schwarz alternating procedure is run on each of the test
vectors for a few iterations with starting guess $0$. Then the initial operators
are constructed from the updated test vectors. This completes the initial
phase. The operators are then updated in the iterative refinement phase
(Alg.~\ref{alg:mgsetup}) that makes use of the full V-cycle of the multigrid
algorithm. For a more detailed description see
\cite{Frommer:2013fsa,Heybrock:lat15}.

\begin{algorithm}[ht]
  \caption{Iterative part of MG setup (standard implementation)}
  \label{alg:mgsetup}
  \SetAlgorithmStyle
  \For {$i=1$ \KwTo $\Nsetup$}
  {
    // apply V-cycle to test vectors\;
    \For {$j=1$ \KwTo $\Ntv$}
    {
      // \emph{coarse-grid correction}\;
      restrict test vector $v_j$ to coarse grid: $v_{c,j} = R\,v_j$\;
      solve coarse system to low accuracy: $u_{c,j} \approx D_c^{-1}\,v_{c,j}$\;
      prolongate result of coarse-grid solve to fine grid: $u_j = P\,u_{c,j}$\;
      // \emph{fine grid}\;
      apply smoother to test vector $v_j$, with result from CGC as starting
      guess\;
      replace test vector $v_j$ by result of smoother\;
    }
    setup of restriction $R$ and coarse-grid operator $D_c$
  }
\end{algorithm}

\section{Basic idea}
After the optimizations described in \cite{Heybrock:lat15}, we identified the
application of the V-cycle to the test vectors in the iterative refinement phase
to be the dominant contribution to the setup time. Therefore our work focuses on
this part of the code exclusively.

In the implementations of \cite{Frommer:2013fsa,Heybrock:lat15} the V-cycle is
applied to the test vectors in a loop sequentially, i.e., to a single right-hand
side (SRHS) at a time. The basic idea of our improvements is simple. We modify
the computation order of the code by blocking the loop over the test vectors
(Alg.~\ref{alg:mgsetup2}) with a block length of $\Nb$ and apply the V-cycle to
multiple right-hand sides (MRHS), i.e., all vectors inside such a block,
simultaneously. Choosing $\Nb = \Nsimd$ and moving the loop inside a block to
the lowest level functions of the code enables us to use this loop for SIMD
vectorization.

\begin{algorithm}[h]
  \caption{Iterative part of MG setup (improved implementation)\protect\footnotemark}
  \label{alg:mgsetup2}
  \SetAlgorithmStyle
  \For {$i=1$ \KwTo $\Nsetup$}
  {
    \For {$j=1$ \KwTo $\Ntv/\Nb$}
    {
      $k = 1 + (j-1)\cdot\Nb,\ \ell = j\cdot\Nb$\;
      apply coarse-grid correction (CGC) to test vectors
      $v_k$,\,\dots,\,$v_\ell$\;
      apply smoother to test vectors $v_k$,\,\dots,\,$v_\ell$, with result from
      CGC as starting guess\;
      replace test vectors $v_k$,\,\dots,\,$v_\ell$ by result of smoother\;
    }
    setup of restriction $R$ and coarse-grid operator $D_c$\;
  }
\end{algorithm}
\footnotetext{%
  In the description of the algorithm we assume that $\Ntv$ is an integer
  multiple of $\Nb$. If it is not the algorithm gets modified in a
  straightforward way, but then part of the SIMD unit is wasted in the last
  iteration, see also \cite{Heybrock:lat15}.
}

\section{Communication bandwidth}

\begin{figure}[ht]
  \centering
  \input{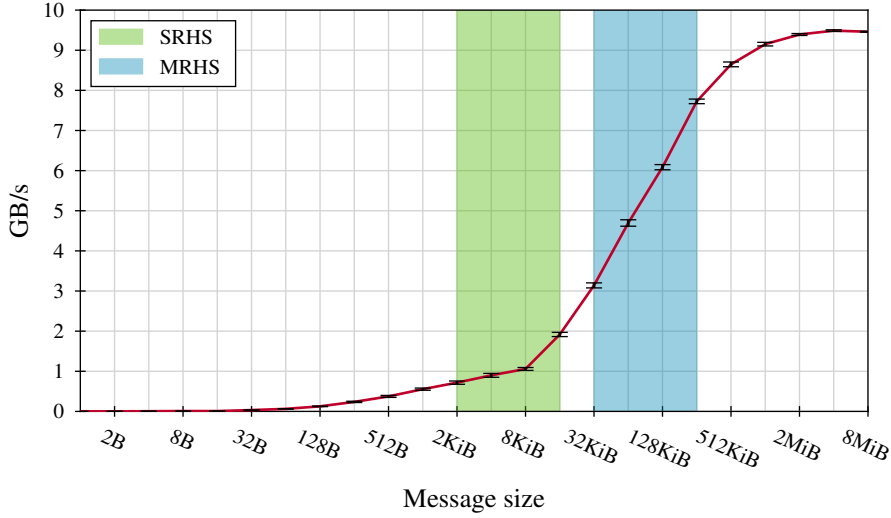}
  \caption{Network bandwidth vs. message size between two KNCs (bi-directional)
  in \qp\ via FDR InfiniBand. Typical message sizes in the \WMG\ setup are shown
  for SRHS (green) and MRHS (blue) setup.}
  \label{plot:comm-bw}
\end{figure}

The effective network bandwidth for off-chip communication via MPI depends on
the message size (cf.~Fig.~\ref{plot:comm-bw}). For small messages latency
effects are dominant, resulting in low bandwidth. For larger messages the
effective bandwidth increases since latency effects become negligible.

The message size ($S_\mu$ in direction $\mu$) on the coarse grid of the \WMG\
setup depends on the local volume of one MPI rank and the degrees of freedom per
site ($2\Ntv$):
\begin{align}
  S_\mu &= \prod_{\nu = 0, \nu \neq \mu}^{3} \frac{\text{(local
  lattice)}_\nu}{\text{(domain size)}_\nu} \cdot \frac{2 \Ntv }{2} \cdot
  \unit[8]{Byte}\,,
  \label{eq:sizes}
\end{align}
where the factor $2$ in the denominator accounts for even-odd preconditioning
and $\unit[8]{Byte}$ is the size of a complex number in single precision. From
Eq.~\eqref{eq:sizes} we obtain message sizes of order 1 KiB for the default
setup, which is just the point where the effective bandwidth starts to increase
significantly. By processing multiple right-hand sides simultaneously we are
able to perform all halo exchanges and global sums, respectively, in the same
call to MPI. Thus the message size increases by a factor of $\Nb = \Nsimd = 16 =
2^4$. Fig.~\ref{plot:comm-bw} suggests an estimated increase in effective
bandwidth of a factor $3\sim4$, which might translate directly to the wall-clock
time spent in communication.

\section{Mapping to SIMD registers}
\label{section:mapping}

The basic layout for data structures based on complex numbers in the original
Wuppertal implementation did not take vectorization into account and uses the
\emph{complex} data type in C. This way, a vector-like object $v$ of length
$\ell$ is stored such that the real and imaginary parts alternate in memory:
\begin{center}
  \begin{tabularx}{.6\textwidth}{|c *{7}{|Y}|}
    \hline &&&&&&\\[-4.5mm]
    $\re v_1$ & $\im v_1$ &
    $\re v_2$ & $\im v_2$ &
    $\cdots$ & $\re v_\ell$ & $\im v_\ell$ \\[.7mm]
    \hline
  \end{tabularx}
\end{center}
This is known as Array-of-Structs (AoS) layout. In the code parts relevant for
us, the implementation in \cite{Heybrock:lat15} works with this layout by
de-interleaving two registers using swizzle intrinsics before and after doing a
SIMD computation. This introduces additional overhead that could be avoided with
a data layout more suitable to vectorization.

Our implementation uses another index for vectorization, i.e., the index of the
different right-hand sides inside a block. For each vector index $i$ we store
$\Nsimd$ $(=16)$ real parts of the right-hand sides followed by the
corresponding imaginary parts:
\begin{center}
  \begin{tabularx}{.8\textwidth}{|c *{8}{|Y}|}
    \hline &&&&&&&\\[-3.5mm]
    $\re v^{(1)}_i$ & $\re v^{(2)}_i$ &
    $\cdots$ & $\re v^{(16)}_i$ &
    $\im v^{(1)}_i$ & $\im v^{(2)}_i$ &
    $\cdots$ & $\im v^{(16)}_i$ \\[1.3mm]
    \hline
  \end{tabularx}
\end{center}
This is known as Array-of-Structs-of-Short-Vectors (AoSoSV) layout. While the
conversion required non-trivial programming effort, this layout yields a more
natural mapping to SIMD. The de-interleaving overhead is gone, and the
individual entries in the registers contain data independent of one another,
which eliminates the need for reduction operations over the elements in the
register.

With our modifications to the data layout, matrix-vector multiplications become
matrix-matrix multiplications, which enables us to use a different vectorization
scheme. In contrast to \cite{Heybrock:lat15} we vectorize the restriction of a
vector from the fine to the coarse grid (as an example) by broadcasting the
elements of the projection operator $R$ as shown in
Alg.~\ref{alg:mrhs_restriction} (see \cite{Heybrock:lat15} for the definition of
$\Nblock$, $V_\text{block}$, and $y_c$). The same vectorization scheme is used
for the application of $D_c$ in the coarse-grid solve. BLAS-like linear
algebra (e.g., vector adds) is vectorized trivially with this data layout.

\medskip

\begin{algorithm}[H]
  \caption{SIMD implementation of restriction $R y = y_c$ with $\Nsimd$
  right-hand sides}
  \label{alg:mrhs_restriction}
  \SetAlgorithmStyle
  \For {$i=1$ \KwTo $\Nblock$}
  {
    \ForEach {$h=\ell,r$}
    {
      \For {$n=1$ \KwTo $6V_{\text{block}}$}
      {
        load real and imag.\ parts of $\Nsimd$ rhs for entry $y_{i,n}^h$ into
        SIMD vectors\;
        \For {$j=1$ \KwTo $\Ntv$}
        {
          load real and imag.\ parts of \smash{$(y_c)^h_{i,j}$} for $\Nsimd$
          rhs\;
          \mbox{broadcast real and imag.\ part of entry $j$ in column $n$ of
          $R_i^h$ into SIMD vectors\hspace*{-10mm}}\;
          increase $(y_c)^h_{i,j}$ by complex fused multiply-add and
          write to memory\;
        }
      }
    }
  }
\end{algorithm}

\section{Memory-bandwidth and cache-reuse considerations}
\label{section:cache}

A dense complex matrix-vector multiplication $c = A \cdot b$, where $A$, $b$,
and $c$ are of dimension $M \times K$, $K$, and $M$, respectively, requires
transferring $(2M + M \cdot K + K)\cdot\unit[8]{Byte}$ from and to memory in
single precision. The computation needs a minimum of $4\cdot M \cdot K/16$
cycles, where a complex \emph{fmadd} consists of 4 real \emph{fmadd} operations,
of which a KNC core can perform $16$ in one cycle. The ratio of these numbers
yields the memory bandwidth per core required to avoid stalls, i.e., $32 \cdot
(2/K + 1 + 1/M)$ Byte/cycle. For a typical working set of a core, $K$ and $M$
are large enough so that their contribution $2/K+1/M$ is negligible compared to
1.\footnote{$M$ needs to be multiplied by $2$ for the spin-splitting of \WMG.}
The resulting required memory bandwidth is then $32$~Byte/cycle per core, or
$2377$~GB/s on $60$ cores of a KNC with a clock speed of $\unit[1.238]{GHz}$,
which is well above the KNC's sustained memory bandwidth of $150-170$~GB/s,
measured with the STREAM benchmark.

Performing the analogous calculation for the matrix-matrix multiplication with
$\Nsimd$ right-hand sides ($A = M \times K$, $B = K \times \Nsimd$, and $C = M
\times \Nsimd$) yields $32 \cdot (2/K + 1/\Nsimd + 1/M)~\text{Byte/cycle}\sim
2~\text{Byte/cycle}=149~\text{GB/s}$. Here, an element of $A$ can stay in cache
for $\Nsimd$ right-hand sides, which results in the difference to the value
above. Thus our method is able to reduce the memory bandwidth requirements of
this code part significantly, and our estimate is now within reach of the KNC's
sustained memory bandwidth.

\section{Results}

At the time of this writing we have finished the implementation of the
coarse-grid solve and the projection operators, while the smoother
\cite{Heybrock:2014iga} still works with the default data layout. This
introduces some temporary copying overhead which will disappear as soon as we
have a MRHS implementation of the smoother.

The results below are from runs on the CLS lattice C101 ($ 48^3\times96$,
$\beta=3.4$, $m_{\pi}=\unit[220]{MeV}$, $a=\unit[0.086]{fm}$) described in
\cite{Bruno:2014jqa}. We use $\Nsimd = 16$ test vectors, a domain size of $4^4$,
and a relative coarse-grid tolerance of $0.05$. The remaining solver
parameters are tuned for minimal propagator wall-clock time with the default
(SRHS) setup. To exclude algorithmic effects and allow for a direct comparison
we use the same parameter combination also for the MRHS setup.
\nopagebreak
\begin{figure}
  \centering
  \input{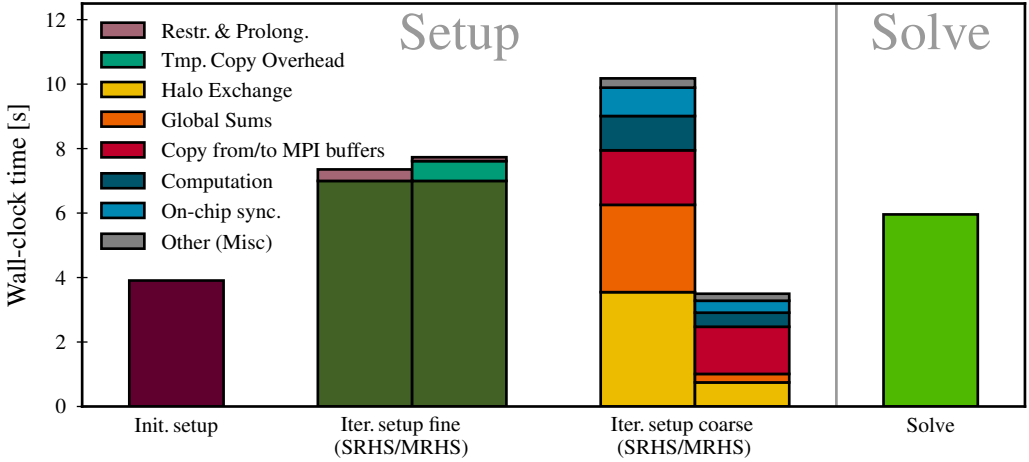}
  \caption{Summary of contributions to the wall-clock time in the \WMG\ setup on
  $64$ KNCs in \qp\ (solve time for comparison). Improvements in parts
  affected by our modifications are given in detail.}
  \label{plot:results}
\end{figure}

With these parameters, a lattice vector on the coarse grid requires
$\unit[1]{\%}$ of the memory of a vector on the fine grid. With the MRHS setup,
we need $32$ lattice vectors on the fine grid and $16 \cdot 32 = 512$ on the
coarse grid. This is to be compared to $17$ and $32$ with the SRHS setup. Thus,
our method needs roughly a factor of $2.2$ more memory in total for the setup.
In a realistic measurement run we typically also keep around several propagators
on the fine grid (consisting of 12 vectors each), so the increase in total
memory consumption is actually considerably smaller.

In Fig.~\ref{plot:results} we show the improvements in wall-clock time we
achieve with our method. We gain a factor of $2.9$ in the projection operators
and a factor of $2.4$ in computation on the coarse grid. Our method needs fewer
calls to barriers between threads, which yields an improvement of $2.7$x in
on-chip synchronization. However, the largest gains are in halo exchanges
($4.7$x) and global sums ($10.3$x), which were the dominant contributions
previously. After our improvements, the wall-clock time is now dominated by
copying data from and to MPI buffers. This is currently done by a single thread
on a single core. In the future we will reduce the impact of these copy
operations by threading them over cores, which will allow us to exploit a larger
fraction of the KNC's sustained memory bandwidth.

In total, the time spent on the coarse grid is reduced by a factor of $2.9$,
which translates to a factor of $1.4$ for the total setup time of \WMG.

\section{Conclusions and outlook}

By combining multiple right-hand sides we were able to significantly reduce the
wall-clock time of the previously dominant contribution (i.e., coarse-grid
solve) to the setup of \WMG, see Fig.~\ref{plot:results} for details. Our
biggest improvements are in communication, where we can send fewer messages that
are larger and thus are able to reduce the impact of latency effects.
Additional improvements were made in computation and on-chip synchronization.

As mentioned above, the impact of the red block in Fig.~\ref{plot:results} (copy
from/to MPI buffers) will be reduced in the future by threading these copy
operations over cores. More importantly, Fig.~\ref{plot:results} shows that the
biggest optimization potential is now in the fine-grid part of the iterative
setup. Therefore we will complete the multiple right-hand-side V-cycle by
applying the techniques used in the present work also to the smoother
\cite{Heybrock:2014iga}, which should yield similar speedups.

\bibliographystyle{jbJHEP_notitle}
\bibliography{Bibliography}

\end{document}